\documentclass[12pt]{article}

\usepackage{epsfig}
\usepackage{epsfig}
\newcommand{\be}{\begin{equation}}
\newcommand{\ee}{\end{equation}}
\newcommand{\bea}{\begin{eqnarray}}
\newcommand{\eea}{\end{eqnarray}}

\begin{document}


\begin{flushright}
LA-UR-03-4509-Rev \\
nucl-th/0309018\\
\today
\end{flushright}

\baselineskip=.33in

\begin{center}

{\large{\bf
Detection of Antineutrinos for Non-Proliferation}}\\ %

Michael Martin Nieto,$^{a}$ A. C. Hayes,$^b$ 
William B. Wilson,$^{c}$ Corinne M. Teeter,$^d$ \\ 
and William D. Stanbro$^{e}$ \\

{\it $^{a,b,c}$Theoretical, $^{d}$Physics, 
and  $^{e}$Nuclear Nonproliferation Divisions, \\  
Los Alamos National Laboratory,\\
University of California, Los Alamos, New Mexico 87545, 
U.S.A.}$^\ddagger$

{\bf Abstract}	

\end{center}

We discuss the feasibility of using the detection of electron 
antineutrinos produced in fission to monitor the time dependence of
the plutonium content of nuclear power reactors.  If practical such 
a scheme would allow world-wide, automated monitoring of reactors 
and, thereby, the detection of certain proliferation scenarios.
For GW$_e$ power reactors the count rates and the sensitivity 
of the antineutrino spectrum (to the core burn-up) 
suggest that monitoring of the gross operational status
of the reactor from outside the containment vessel is feasible.
As the plutonium content builds up in a given burn cycle the total 
number of antineutrinos steadily drops and this variation is quite
detectable, assuming fixed reactor power.  The average antineutrino 
energy also steadily drops, and a measurement of this variation would 
be very useful to help off set uncertainties in the total reactor 
power.  However, the expected change in the antineutrino signal from 
the diversion of a significant quantity (SQ) of plutonium, which
would typically require the diversion of as little as a single fuel
assemblies in a GW$_e$ reactor, would be very difficult to detect.

\vspace{.75in}
\baselineskip=.165in

\noindent\underline{~~~~~~~~~~~~~~~~~~~~~~~~~~}\\
$^\ddagger${\small{Mail stops and email addresses: 
$^{a}$B285, mmn@lanl.gov;  
$^{b}$B227, anna\underline{~}hayes@lanl.gov; 
$^{c}$B243, wbw@lanl.gov;
$^{d}$D454, cteeter@lanl.gov;
$^{e}$E541, wstanbro@lanl.gov
}}


\newpage
\baselineskip=.33in


\section{Background}

It is widely reported that India obtained its weapons-grade plutonium by
running its un-safeguarded CANDU reactors to produce Pu 
\cite{candu1}--\cite{candu3}.  Prevention of
the illicit production of Pu in most countries is through the
imposition of an intrusive inspection regime by the International
Atomic Energy Agency. This approach has proven to be quite resource
intensive.  More preferable would be a method to monitor {\it continually} the
nuclear fuel content of reactors, thus ensuring that weapons material
is not being diverted.  Ideally, one wants to do this not only
continuously, but also cheaply, unobtrusively, and with minimal manpower. 

Related (but somewhat independent) concerns are the difficulty of 
determining the isotopic fuel content, for attribution,  of a rogue
nuclear device and the difficulty of detecting
a small nuclear explosion of order 1 kton.  Such a small explosion,
done underground, can be concealed because of the lack of any definitive 
signal; for example, microseisems would mask any acoustic signature.   

For all these problems electron antineutrino\footnote{
The term ``neutrino'' is often generically used to
describe both neutrinos and their antiparticle antineutrinos, especially
with respect to a ``neutrino detector.''  We will do the same for a
detector or beam, but will try to make the distinction clear for
particle processes.}
detection in principle offers a solution.  The feasibility of
using antineutrino detection has been examined in the past.
In fact, the original reactor experiment which
discovered the antineutrino (by Fred Reines and Clyde Cowan) was done  
at a reactor only after a first idea had been 
considered, to detect antineutrinos from a nuclear explosion at the Nevada
Test Site \cite{nevada}-\cite{nevada3}.

Indeed, the technologies developed for particle physics experiments 
that use large neutrino detectors have added further interest in such a
solution. For example, in recent times large 
water- \cite{superk}-\cite{newSNO}, ice-  \cite{amanda}-\cite{ICECUBE}, 
and mineral-oil-based \cite{LSND,miniboone} 
neutrino detectors have become practical and relatively efficient.  
In fact, Supernova 1987A was seen by  the Kamiokande detector 
in Japan \cite{japan1987} and the IMB detector in the USA  \cite{usa1987}. 
Each detecting about 10 events, as shown together in
Ref. \cite{1987A}.

There have been large and significant experiments on the antineutrino
spectra from reactors \cite{3percent,2percent}.  
Also a well-known, ongoing neutrino oscillation  
experiment detects reactors from distances on the order of 
100 km \cite{kamland}.  The idea of using a mobile nuclear submarine 
to do a similar experiment has been discussed \cite{submarine} and 
the detection of radioactive antineutrinos from the Earth's core is
now feasible \cite{nuearth}.  

Further, at present there already is an ongoing experimental
study \cite{bernstein} on the feasibility of building an 
antineutrino detector for reactor monitoring, a subject with its own
history \cite{ru2,ru1}.   The aim is to 
determine the class of safeguards problems that might be addressed. 
In this paper we examine some of the theoretical issues involved 
in trying to monitor reactor core fuel with antineutrinos.


\section{Physics motivation}
\label{sec2}

When a nucleus undergoes fission the unstable
fission products pre-dominantly beta-decay, thus emitting
antineutrinos.  
Fission processes mostly emit antineutrinos (as opposed
to neutrinos) because the vast majority of the
beta decays are from neutron rich nuclei where the underlying process
is the conversion of a neutron into a proton, electron, and antineutrino:
\be 
n \rightarrow p + e^- +  \bar{\nu}_e. \label{betadecay}
\ee

\begin{figure}[h!]
 \begin{center}
\noindent    
\psfig{figure=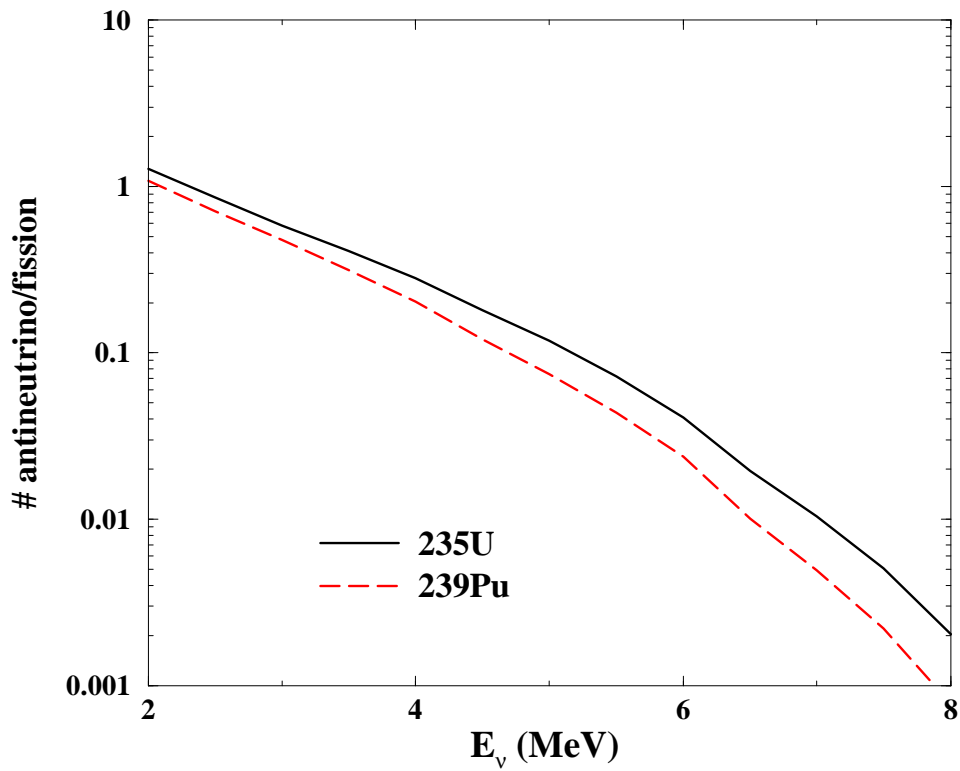,width=5in}
\end{center}
  \caption{The relative energy spectra of the 
antineutrinos from the fission of
 $^{235}$U (dashed line) and  $^{239}$Pu (solid line).  
 \label{unup}}
\end{figure}

On average, about 5 antineutrinos are emitted per fission.
These antineutrinos are not emitted instantaneously because of the
finite life-time of the fission products. 
However, the fact that reactor monitoring is a steady-state 
measurement means that time variation is not an issue.   The energy
spectrum of the antineutrinos ranges from zero to about 15 MeV, with
an average energy of  about 3 MeV. Only a very small fraction of the
antineutrinos emitted have energies above 8 MeV.  As an example, in
Figure \ref{unup} we show the relative energy spectra of the 
antineutrinos from the fission of $^{235}$U and  $^{239}$Pu.


This and the other fission spectra used in this paper were calculated 
assuming thermal neutron induced fission for $^{235}$U, $^{239,241}$Pu and
fast neutron induced fission for $^{238}$U. 
To obtain these results we used the England and
Rider \cite{tal} evaluated cumulative fission 
product yields and the beta decay branching ratios and end-point
energies from ENDF/B-VI \cite{ENDF}.  

The cumulative spectrum for a given nuclide involves contributions
from thousands of beta decays, each associated with an unique end point
energy.  This cumulative spectrum is given by  
\be
N(E_{\bar{\nu}}) = \sum_i
Y_i(A,Z)\sum_{i,j}B_{i,j}(E_0^j)R(E_{\bar{\nu}},E_0^j,Z),
\ee
where the $Y_i(A,Z)$ are the fission product cumulative yields, $B_{i,j}$ are 
the $j$ branching fractions for the decays of nucleus (A,Z) 
with end-point energies $E_0^j$, each described by an individual  
beta decay spectrum $R(E_{\bar{\nu}},E_0^j,Z)$.  

The differences in the fission products produced in the differing fissioning
systems leads to a significant difference in the magnitude and shape
of the respective antineutrino spectra. 
These differences, and the fact that the emitted antineutrinos
cannot be shielded, are key to the concept of monitoring the core fuel.

Antineutrino detection occurs mainly
through the antineutrino-proton charge-current reaction\footnote{
Contrariwise, neutrino detection is mainly via the opposite reaction,
\be
\nu_e + n \rightarrow   p   + e^-.   \label{nun}
\ee 
since neutrinos tend to result from a fusion process, such as in the Sun.}
\be
\bar{\nu}_e + p \rightarrow e^+ + n.  \label{anup}
\ee
The threshold for this reaction is 1.8 MeV and
the cross section at the mean fission 
antineutrino energy is about $10^{-42}$ cm$^2$.
When folded with the energy-dependent detection cross section (\ref{anup})
the detected antineutrino spectrum peaks at about 3.8 MeV.
Because the interaction of neutrinos with matter is so extremely weak, 
the detector sizes that would be needed for projects of the type we
are considering need to be either very large or able to have long
measurement times, or both.\footnote{Antineutrinos can also
interact with the electrons in the detector via antineutrino-electron
scattering.   But the event rate is orders of magnitude smaller
\cite{jbn}, essentially because the mass factor in 
the cross-section equation is that of the electron \cite{emass}.}


\section{The principles of antineutrino monitoring}

We now turn to the feasibility of detecting the
antineutrinos produced in fission to monitor fuel content 
of a functioning nuclear reactor \cite{aboutreactors}, from 
outside the containment building.  
There are several classes of power reactors that need to be considered
\cite{powerreact,powerreact2}.   But in the present work we restrict
our discussion to a Pressurized Water Reactor (PWR) similar to 
the San Onofrie\footnote{Throughout this paper we distinguish the 
thermal from the electric power of a 
reactor using the subscripts $t$ and $e$, respectively. For example,
GW$_t$ refers to a GW of thermal power, while GW$_e$ refers to a GW of
electric power.  Typically, GW$_e \approx 0.3$ GW$_t$.}
3.4 GW$_t$ reactor in California, where  
antineutrino monitoring is presently being studied\cite{bernstein}
(see below). 
We examine the predicted time-dependent antineutrino spectra for
a 2.7\% enriched PWR reactor. To test the sensitivity of the expected
signals to the initial uranium enrichment we also examine
a 4.2\% enriched PWR.

A 3.4 GW$_t$ power reactor emits about $10^{26}$ antineutrinos
per day.  The absolute magnitude of the detected antineutrino spectrum
differs for each of the actinides contributing to the total fission
rate.  So, the total number of antineutrinos detected
changes with the relative fissioning fraction of the these isotopes in
the reactor core.  Thus, for a fixed reactor power, the number of
detected antineutrinos is a reflection of the core burn-up. 

\begin{figure}[h!]
 \begin{center}
\noindent    
\psfig{figure=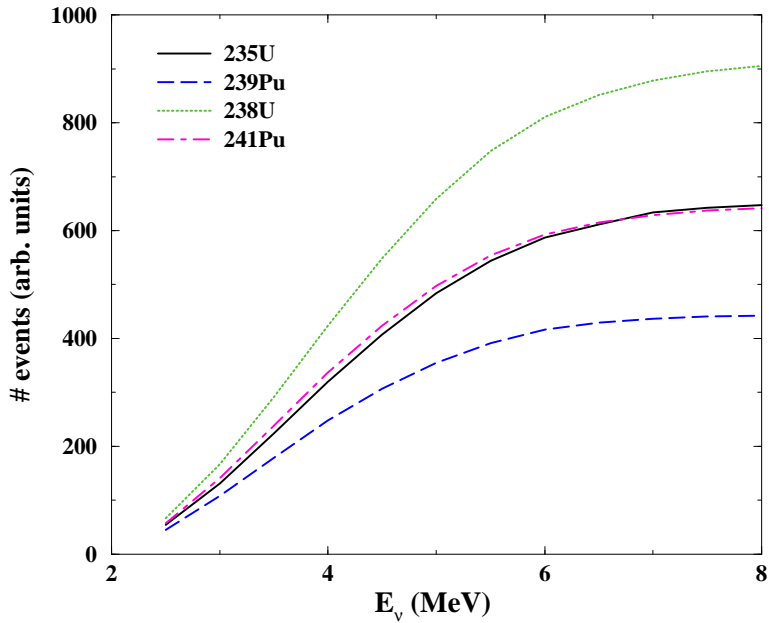,width=4in}
\end{center}
  \caption{The integrated fission antineutrino spectra up to energy $E_\nu$ 
 \label{uintegratepu}}
\end{figure}

The shape of the spectrum is also a measure of the core burn-up.
As shown in Figure  \ref{uintegratepu}, the cumulative number of
antineutrinos (when folded over the detection cross section) as a 
function of antineutrino energy is different for each species.  By
comparing the number of antineutrinos with energies up to 3 MeV
with the number  up to 6 MeV, 
pure $^{235}$U and $^{239}$Pu are easily distinguishable.  

The variation in the number of antineutrinos emitted with burn-up could
be masked by uncertainties in the total thermal reactor power (total
fission rate). 
Thus, if feasible, a measurement of the time-dependent shape of the
spectrum or of the average 
antineutrino energy would provide an important independent
variable for monitoring the reactor core.


A type of detector that could be used would be a portable version
of the mineral-oil based liquid scintillator used in, for example, the
100 ton LSND  neutrino oscillation experiment \cite{LSND} and the
follow-up MiniBooNE experiment \cite{miniboone}.
The Sandia/Livermore collaboration \cite{bernstein} is currently testing 
a prototype neutrino detector of this type.  For a 1 cubic meter
liquid scintillator detector situated 24 m from the core of a 3.4
GW$_t$ reactor, Bernstein et al. \cite{bernstein} 
expect a healthy signal of 2740 antineutrino events per day.

The questions that we attempt to address here are:
\begin{itemize} 
\item  What are the expected isotopic contributions to the fission rate 
as a function of time?
\item what are the corresponding
antineutrino spectra and how well do they reflect the fuel burn-up?
\item  what are the expected changes in the antineutrino spectra
for different fuel diversion scenarios? 
\item and what detection accuracy is required to safeguard against these?   
\end{itemize}

To address these questions we examined a 2.7\% and a 4.2\% enriched uranium
PWR. The 2.7\% PWR corresponds to a 3 year discharge neutron exposure
of 30 GW days per metric ton of uranium (GWd/MTU), whereas, the 4.2\%
enriched fuel 
corresponds to a neutron exposure of twice that in its lifetime. 
The temporal results were taken from earlier coupled calculations with
the ERPI/CELL and CINDER-2 codes \cite{wilsonagain,wilsonNRC}.


\begin{figure}[h!]
 \begin{center}
\noindent    
\psfig{figure=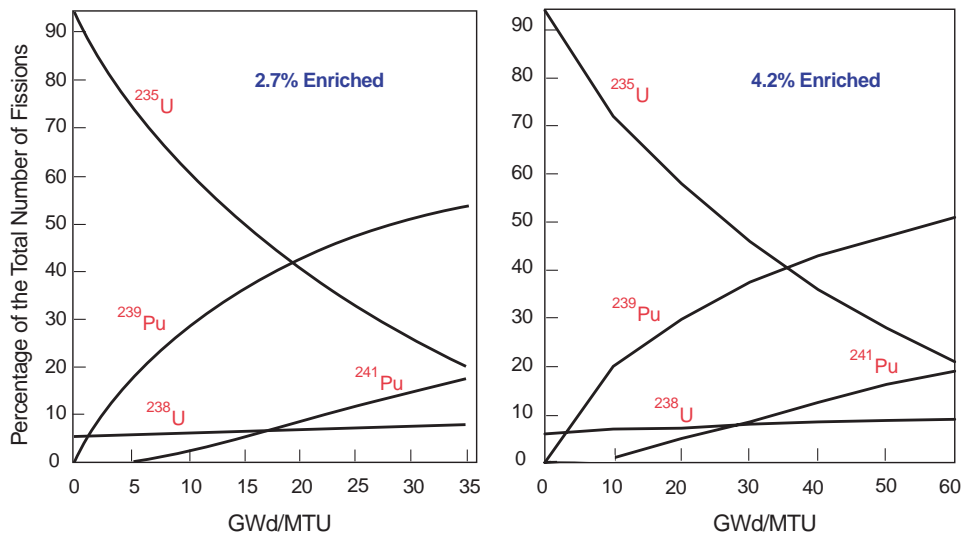,width=5in}
\end{center}
  \caption{The change in fuel composition with time of 
four isotopes in a 2.7\% and 4.2\% pressurized light water reactor.  
 \label{fuelchange}}
\end{figure} 


In Figure \ref{fuelchange} we show the fission history 
for the 2.7\% enriched (97.3\% $^{238}$U, 2.7\% $^{235}$U)  and  the
4.2\%  enriched (95.8\% $^{238}$U, 4.2\% $^{235}$U) reactors.
As can be seen, the percentage of the total fission
from $^{235}$U steadily drops as a  
function of time while that from $^{239}$Pu steadily increases. 
Apart from the factor of two difference in the exposure times, the
fission histories for the two different enrichments are very similar.

The results of all our other calculations for the antineutrino spectra 
for the two different fuel enrichments are also very similar. 
Therefore, for the remainder of this paper we will show only those
results obtained for the 2.7\% enriched case. So, for example, 
our conclusions on antineutrino monitoring apply equally well to the
4.2\% enriched fuel, with the understanding that exposure times should
be approximately doubled. 

After about 3 years burning (30 GWd/MTU) 
the  fuel has an actinide isotopic  composition of 
(95\% $^{238}$U, 1\% $^{235}$U, 1\% $^{239}$Pu, 3\% heavier actinides 
including $^{240,241}$Pu), with the fission fuel contributing to the total
fraction of fissions as shown in Fig. \ref{fuelchange}.

\begin{figure}[h!]
 \begin{center}
\noindent    
\psfig{figure=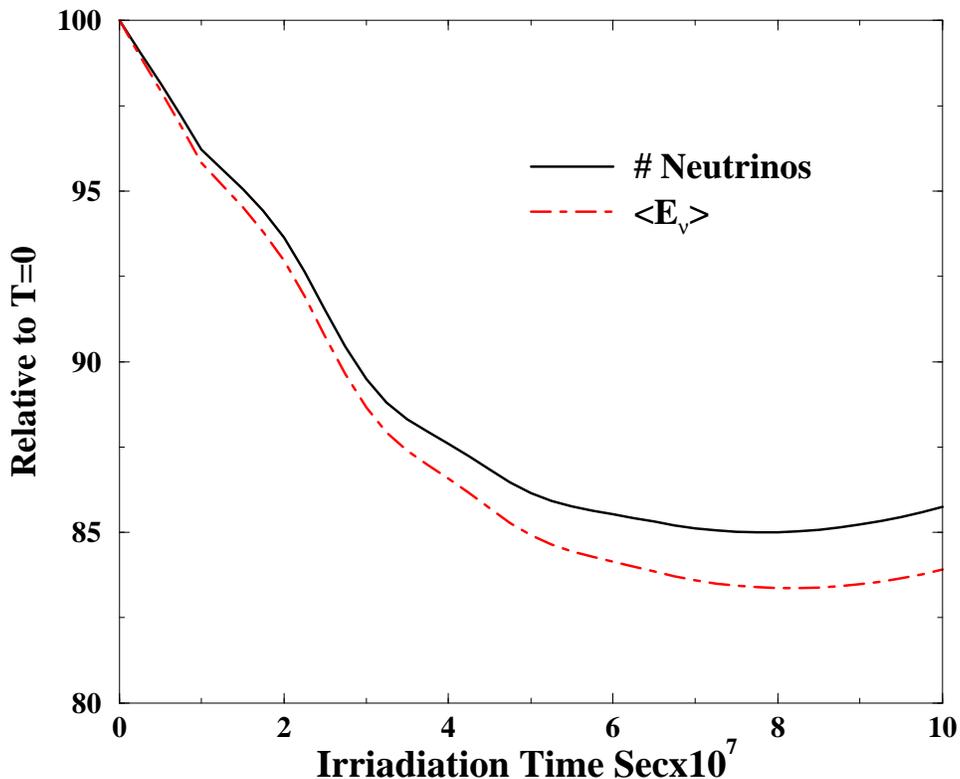,width=5in}
\end{center}
  \caption{The expected change in the number of antineutrino events
and the average antineutrino energy 
as a function of time. The drop in the number of antineutrino events occurs
because, as the fuel burns, the fraction of fissions from $^{239}$Pu
and the number of antineutrinos emitted from $^{239}$Pu is lower than
for $^{235}$U. 
As $^{241}$Pu grows in the number of events levels off. 
 \label{freshfuel}}
\end{figure}

As the $^{239}$Pu accumulates in the reactor, it also captures
neutrons to produce $^{240}$Pu, which in turn captures neutrons to
produce $^{241}$Pu.  The optimum time for clandestine extraction
would therefore depend on the plutonium isotopic ratios desired.  
The maximum plutonium content is after 3 years exposure.  
However, $^{240}$Pu has some undesirable nuclear properties from a weapons
point of view, and better grade plutonium is obtained by removing
the plutonium before a cycle is completed. 
Over the fuel life-time the steady increase in the fission fraction from
$^{239}$Pu translates into a steady decease in the number of
antineutrinos emitted and in the
average antineutrino energy, Fig {\ref{freshfuel}}.
The spectrum of emitted antineutrinos by 
fissioning $^{241}$Pu does not differ significantly from $^{235}$U. 
Thus, after about 1.5 years the number of antineutrino emitted begins
to level off as the $^{241}$Pu starts to build up. 


\section{Monitoring fuel content of a running reactor}
\label{realreactor}

The situation described in Fig {\ref{freshfuel}}
is for fresh enriched uranium irradiated over a  
three year period. This differs from the actual situation
for a normal PWR burning in equilibrium, where the fuel is cycled annually.
A large pressurized light water reactor has about 240 
fuel assemblies, of which 80 are replaced at the end of each cycle,
and the assemblies are then shuffled in their location. 
When the reactor is running in equilibrium at the beginning of each
fuel cycle, one third of the fuel is fresh enriched uranium, 
one-third has been irradiated for one cycle 
and one-third for two cycles. At the end of the cycle one-third of the fuel
has been irradiated for three cycles and is removed and replaced with
fresh fuel.  Each fuel assembly typically contains one-half ton of fuel.
Therefore, in three years about 5 kg of plutonium will be produced in each 
assembly of normal fuel.

As the reactor operates in equilibrium, from the beginning of a cycle
to the end of a cycle. 
the change in the antineutrino spectrum is
a steady (but less steep) drop in the number of antineutrinos 
emitted as  the $^{239}$Pu builds up. For  a 2.7\% enriched uranium
PWR the drop in the number of antineutrinos is about 5\% in one
cycle. When the discharged fuel is replaced  
with fresh fuel, the antineutrino event rate jumps to its `start
of cycle' value.  However, it is important to note that monitoring the
reactor core fuel burn-up through the time-dependent change in the
antineutrino spectrum requires independent knowledge of the reactor power.
Otherwise the expected 5\% change in the antineutrino spectrum might 
be masked by unrecorded fluctuations
in the total power.

Typical diversion scenarios would likely involve the diversion 
of an entire fuel assembly and replacement with fresh fuel. 
In principle, a single assembly would contain enough plutonium to make
a nuclear weapon.  An unannounced removal of 1-2 additional fuel
assemblies at the same time that other scheduled work 
was being carried out would require an accuracy in the antineutrino detection
rate of better than 1\%.
The detection of this type of change would be a 
much more difficult task than observing the 5\% change in the magnitude of
the spectrum over one fuel cycle.


\begin{figure}[h!]
 \begin{center}
\noindent    
\psfig{figure=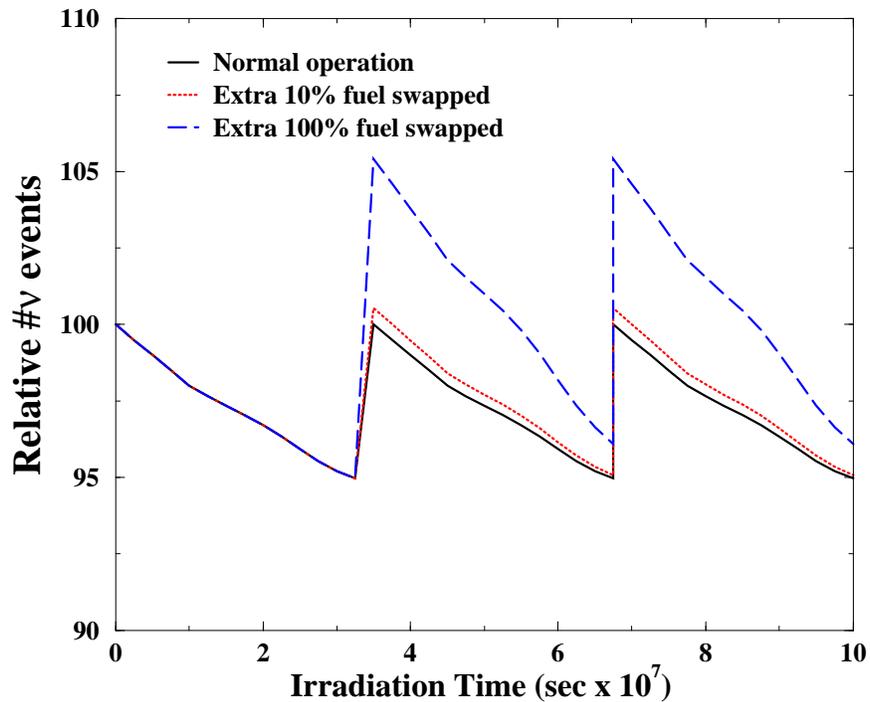,width=4.5in}
\end{center}
  \caption{When burning in equilibrium the number of antineutrino
events drops about 5\% from the beginning to the end of a normal fuel
cycle in a PWR (solid line).  . 
A diversion of  an additional 10\% of the fuel (more
than a critical mass of Pu) leads to a small
change in the number of antineutrinos emitted (dotted line). The gross
misuse of the reactor by the replacement 
of an addition 1/3 of the irradiated fuel with fresh fuel during
fuel-cycle management (long-dashed curve) 
would lead to an observable change in the antineutrino spectrum.} 
 \label{equilfuel}
\end{figure} 



In Fig. \ref{equilfuel} we compare the expected time-dependent
antineutrino count rate under normal operation of a 2.7\% PWR with that
expected for a significant fuel diversion. In this latter case we
have assumed that an additional unannounced 10\% of the fuel was replaced
during a scheduled fuel cycle management. The solid curve shows the relative 
change in the number of events on normal fuel management, i.e., at the end of
the fuel cycle the discharged fuel is removed and at the start of the
next cycle the core  
consists of 1/3 fresh 2.7\% uranium fuel, 1/3 fuel irradiated for 1 cycle, 
and 1/3 irradiated for 2 cycles.
The dotted curve shows a realistic and significant violation of their 
safeguards obligations in which an additional unreported
10\% of the 2 cycle-irradiated fuel has been replace with fresh fuel. 
In this case the start of the new cycle
involves a core with 37\% fresh 2.7\% enriched uranium, 
33\% 1 cycle irradiated fuel, and 30\% 2 cycle irradiated fuel.
We note, that the typical shutdown time of 1 month during fuel management 
is not included in Fig. \ref{equilfuel}.

In contrast to the above 10\% diversion of the fuel scenario, gross
diversions of fuel or gross 
deviations in the reactor operation from that announced would likely
lead to quite detectable 
changes from the expected antineutrino spectrum. 
The long-dashed curve in Fig. \ref{equilfuel}
represents the expected number of antineutrino in the case that 2/3
(as opposed to the regular 1/3) of the irradiated fuel is replaced at
the end of a cycle.  In this scenario, at the start of the new cycle
the core would consist of 2/3 fresh 2.7\% uranium and 1/3 1 cycle
irradiated fuel.  This gross misuse of the reactor would lead to a
10\% shift in the antineutrino count rate, to be compared with the 5\%
change expected under normal fuel-cycle management. 

All the calculations presented above for the expected changes in
antineutrino spectra assume that our reactor burn-up and antineutrino
spectra calculations assume no uncertainties.  This, of course, is
not the case. Uncertainties in the calculated inventories of the four
important fissioning materials ($^{235,238}$U, $^{239,241}$Pu) 
compared with the 
inventories measured \cite{wilsonNRC} for a range of fuels indicate
that these inventory calculations are accurate to better than a few
percent.  

However, the inventories of total core radionuclides is of
limited extent and the uncertainties large. 
This precludes a determination of the uncertainties on inventories of
the important fission products that 
contribute to the aggregate beta and antineutrino spectrum. 
For these, inventories calculated by CINDER and ORIGEN can deviate by
more than 10\% in several cases.  The accuracy with which reactor core
burn-up and the corresponding antineutrino spectrum can be monitored
is best determined by a comparison between calculated expectations and
experimental observations of the time-dependent change antineutrino
spectrum for a known reactor. 

The ongoing work at the San Onofrie
reactor \cite{bernstein} could provide ideal 
data for a comparison between theory and experiment.  
However, even though large reactor detectors for neutrino oscillation
experiments, such as CHOOZ \cite{3percent}, have measured the energy
dependence of the antineutrino spectrum quite well,  
small detectors of the San Onofrie size ($\sim 1$ m$^3$) will be
unlikely to provide the desired information on the antineutrino
spectral shape.  Since, as discussed, the difference in the detected
antineutrino energy spectral shape for $^{235}$U and $^{239}$Pu is
quite significant, such information would be very 
important in monitoring the fuel content of a reactor.  

For smaller reactors ($\sim$ 10 MW)  a greater fraction of the fuel
would need to be diverted at one time.  In assessing the feasibility of using
antineutrino detection for such reactors there is a trade off between
the  smaller number of fissions per day (and hence the number of
detectable antineutrinos) and the larger change in the spectrum.  
Antineutrino detection also presents itself as
a possibly useful means of monitoring hot spent fuel \cite{divert}.


\section{conclusions}

Antineutrino monitoring of reactors can provide unique information on
the burn-up in the core from outside the containment vessel.
If accurate knowledge of the reactor power is known through an
independent measurement, the variation of the number of detected
antineutrino events reflects the build-up of plutonium.  Thus,
antineutrino monitoring could be used to detect gross deviations from
the declared operational mode of a reactor. 
A measurement of the average anti-neutrino energy or of the shape of
the spectrum would provide valuable additional information and would
greatly reduce uncertainties in relating the antineutrino spectrum to
core burn-up. However, the magnitude of the expected change in the
antineutrino count rate of less than 1\% in the case of the diversion
of a critical mass of plutonium suggests that antineutrino 
monitoring is unlikely to be sensitive to this class of safeguards.

\section*{Acknowledgments}

We thank Adam Bernstein, James E. Cahalan, Stacey Eaton, Jim Friar,  
Giorgio Gratta, Tom Kunkle, 
John Learned, Chris Morris, Vern Sandberg, and Steve Sterbenz 
for helpful comments and  questions.  
This work was supported by the United States Department of Energy. 



\end{document}